%% file: RR-6524.tex
\pdfoutput=1
\documentclass[11pt]{article}
\usepackage{RR}
\RRNo{6524}

\usepackage{latexsym,amssymb}
\usepackage{graphicx,epsfig}
\usepackage{theorem,subfigure}
\usepackage{algorithm}

\floatstyle{ruled}
\newfloat{Algorithm}{htbp}{lop}
\newfloat{Schedule}{htbp}{lop}

\newtheorem{theorem}{Theorem}
\newtheorem{lemma}{Lemma}
\newtheorem{definition}{Definition}

\newcommand{\A}{\mathtt{A}}

\newcommand{\M}{\mathtt{M}}

\newcommand{\mod}{\mathrm{mod}}

\newenvironment{proof}{\noindent\textbf{Proof:}}{\hfill$\Box$}

\RRdate{May 2008}
\RRtitle{Bornes pour l'auto-stabilisation dans les r\'{e}seaux unidirectionnels}
\RRetitle{Bounds for self-stabilization in unidirectional networks}
\titlehead{Bounds for self-stabilization in unidirectional networks}
\RRauthor{Samuel Bernard$^\star$ \and St\'{e}phane Devismes$^\circ$ \and Maria Gradinariu Potop-Butucaru$^{\star,\dag}$ \and S\'{e}bastien Tixeuil$^{\star,\ddag}$\\
{\small
$^\star$ Universit\'{e} Pierre et Marie Curie - Paris 6, LIP6, France\\
$^\circ$ CNRS, LRI, France\\
$^\dag$ INRIA project-team Regal\\
$^\ddag$ INRIA project-team Grand Large
}
}

\authorhead{S. Bernard \emph{et al.}}

\RRresume{
Un algorithme r\'eparti est auto-stabilisant si, apr\`es que des fautes et des
attaques aient frapp\'e le syst\`eme et l'aient plac\'e dans un \'etat
quelconque, le syst\`eme corrige cette situation catastrophique sans
intervention ext\'erieure en temps fini. Les r\'eseaux unidirectionels
emp\^echent de nombreuses techniques habituelles dans le cadre de
l'auto-stabilisation d'\^etre utilis\'ees, comme la pr\'{e}servation de
pr\'{e}dicats locaux. Dans cet article, nous \'evaluons la complexit\'e
intrins\`eque de la r\'ealisation de l'auto-stabilisation dans les r\'eseaux
unidirectionels, et nous nous concentrons sur le probl\`eme du coloriage de
n\oe uds.

Quand les solutions d\'eterministes sont envisag\'ees, nous prouvons qu'il
existe une borne inf\'erieure de $n$ \'etats par processus (o\`u $n$ est la
taille du r\'eseau) et un temps minimal avant retour \`a la normale de
$n(n-1)/2$ action au total. Nous pr\'esentons un algorithme d\'eterministe dans
des r\'eseaux de topologies quelconques qui est optimal en espace et en temps.

Dans le cadre de solutions probabilistes, nous observons que $\Delta+1$ \'etats
par processus et $\Omega(n)$ actions au total sont requises (o\`u $\Delta$
repr\'esente le degr\'e du graphe non-orient\'e sous-jacent). Nous pr\'esentons
un algorithme auto-stabilisant probabiliste qui utilise $\mathtt{k}$ \'etats
par processus, o\`u $\mathtt{k}$ est un param\`etre de l'algorithme. Quand
$\mathtt{k}=\Delta+1$, l'algorithme retrouve un comportement correct en
$O(\Delta n)$ actions en moyenne. Mais si $\mathtt{k}$ augmente de mani\`{e}re
arbitraire, l'algorithme retrouve un comportement correct en $O(n)$ actions en
moyenne. Au final, notre algorithme peut \^etre rendu optimal en espace ou en
temps. 
}

\RRabstract{
A distributed algorithm is self-stabilizing if after faults
and attacks hit the system and place it in some arbitrary global state, the
systems recovers from this catastrophic situation without external intervention
in finite time. Unidirectional networks preclude many common techniques in
self-stabilization from being used, such as preserving local predicates. In
this paper, we investigate the intrinsic complexity of achieving
self-stabilization in unidirectional networks, and focus on the classical 
vertex coloring problem. 

When deterministic solutions are considered, we prove a lower bound of $n$
states per process (where $n$ is the network size) and a recovery time of at
least $n(n-1)/2$ actions in total. We present a deterministic algorithm with
matching upper bounds that performs in arbitrary graphs. When probabilistic
solutions are considered, we observe that at least $\Delta + 1$ states per
process and a recovery time of $\Omega(n)$ actions in total are required (where
$\Delta$ denotes the maximal degree of the underlying simple undirected
graph). We present a probabilistically self-stabilizing algorithm that uses
$\mathtt{k}$ states per process, where $\mathtt{k}$ is a parameter of the
algorithm. When $\mathtt{k}=\Delta+1$, the algorithm recovers in expected
$O(\Delta n)$ actions. When $\mathtt{k}$ may grow arbitrarily, the algorithm
recovers in expected $O(n)$ actions in total.  Thus, our algorithm can be made
optimal with respect to space or time complexity.
}

\RRmotcle{auto-stabilisation, bornes inf\'erieures, r\'eseaux unidirectionels, coloriage
}
\RRkeyword{self-stabilization, lower bounds, unidirectional networks, coloring
}
\RRprojet{Grand Large}
\RRtheme{\THNum}
\URFuturs

\begin{document}
\makeRR

\input{intro}
\input{model}
\input{impossibility}

\input{lower_bounds}
\input{deterministic}
\input{probabilistic}
\input{conclusion}

\bibliographystyle{plain}
\bibliography{../../../biblio/biblio}

\end{document}

%% file: intro.tex
\section{Introduction}

One of the most versatile technique to ensure forward recovery of distributed
systems is that of \emph{self-stabilization}~\cite{D74j,D00b}. A distributed
algorithm is self-stabilizing if after faults and attacks hit the system and
place it in some arbitrary global state, the systems recovers from this
catastrophic situation without external (\emph{e.g.} human) intervention in
finite time. Self-stabilization makes no hypotheses about the extent or the
nature of the faults and attacks that may harm the system, yet may induce some
overhead (\emph{e.g.} memory, time) when there are no faults, compared to a
classical (\emph{i.e.} non-stabilizing) solution. Computing space and time
bounds for particular problems in a self-stabilizing setting is thus crucial to
evaluate the impact of adding forward recovery properties to the system.

The vast majority of self-stabilizing solutions in the literature~\cite{D00b}
considers bidirectional communications capabilities, \emph{i.e.} if a process
$u$ is able to send information to another process $v$, then $v$ is always able
to send information back to $u$. This assumption is valid in many cases, but
can not capture the fact that asymmetric situations may occur, \emph{e.g.} in
wireless networks, it is possible that $u$ is able to send information to $v$
yet $v$ can not send any information back to $u$ ($u$ may have a wider range
antenna than $v$). Asymmetric situations, that we denote in the following under
the term of \emph{unidirectional} networks, preclude many common techniques in
self-stabilization from being used, such as preserving local predicates (a
process $u$ may take an action that violates a predicate involving its outgoing
neighbors without $u$ knowing it, since $u$ can not get any input from them).

\paragraph{Related works}

Self-stabilization in bidirectional networks makes a distinction between
\emph{global} tasks (\emph{i.e.} tasks whose specification forbids particular
state combinations of processes arbitrarily far from one another in the
network, such as leader election) and \emph{local} tasks (whose specifications
forbid particular state combinations only for processes that are at distance at
most $d$ from one another, for some parameter $d$). Local tasks are often
considered easier in bidirectional networks since detecting incorrect
situations requires less memory and computing power~\cite{BDDT07j}, recovering
can be done locally~\cite{AD02j}, and Byzantine containment can be
guaranteed~\cite{MT07j,NA02c}.

Since a self-stabilizing algorithm may start from any arbitrary state, lower
bounds for non-stabilizing (\emph{a.k.a.} properly initialized) distributed
algorithms still hold for self-stabilizing ones. As a result, relatively few
works investigate lower bounds that are specific to
self-stabilization~\cite{BGJ07j,DGS99j,DIM97j,DHT04ca,GT02c,T01c}. Results
related to space lower bounds deal with global tasks (\emph{e.g.} constructing
a spanning tree~\cite{DGS99j}, finding a center~\cite{DGS99j}, electing a
leader~\cite{BGJ07j,DGS99j}, passing a token~\cite{DIM97j,DHT04ca,T01c}, etc.).
\cite{GT02c} provides a time lower bound for self-stabilizing token passing,
still a global task. Global tasks typically require $\Omega(n)$ states per
process (\emph{i.e.} $\Omega(\log(n)$ bits per process) and $\Omega(n)$ time
complexity to recover from faults.

Investigating the possibility of self-stabilization in unidirectional networks
was recently emphasized in several
papers~\cite{AB98j,CG01c,DDT99j,DDT06j,DS04j,DT01jb,DT03j}\footnote{We do
consider here the overwhelming number of contributions that assume a
unidirectional ring shaped network, please refer to~\cite{D00b} for additional
references}. In particular, \cite{DDT99j} show that in the simple case of
acyclic unidirectional networks, nearly any recursive function can be computed
anonymously in a self-stabilizing way. Computing global tasks in a general
topology requires either unique identifiers~\cite{AB98j,CG01c,DS04j} or
distinguished processes~\cite{DDT06j,DT01jb,DT03j}. Observe that all
aforementioned works consider global tasks, and provide constructive upper
bound results (\emph{i.e} algorithms), leaving the question of matching lower
bounds open.

\paragraph{Our contribution}

In this paper, we investigate the intrinsic complexity of achieving
self-stabilization in unidirectional networks, and focus on the classical
vertex coloring problem, a local task with several known efficient
self-stabilizing solutions in bidirectional networks~\cite{GT00c,MFGST06c}.
Deterministic and probabilistic solutions require only a number of states that
is proportional to the network maximum degree $\Delta$, and the number of
actions per process in order to recover is $O(\Delta)$ (in the case of a
deterministic algorithm) or expected $O(1)$ (in the case of a probabilistic
one). To satisfy the vertex coloring specification in unidirectional networks,
an algorithm must ensure that no two neighboring nodes (\emph{i.e.} two nodes
$u$ and $v$ such that either $u$ can send information to $v$, or $v$ can send
information to $u$, but not necessarily both) have identical colors. 

The main result of this paper is to show that solving a \emph{local} task in
unidirectional networks with a deterministic algorithm is as difficult as
solving a \emph{global} task in bidirectional networks, while nice complexity
guarantees can be preserved with probabilistic solutions: 
\begin{enumerate}
\item When deterministic solutions are considered, we prove a lower bound of
$n$ states per process (where $n$ is the network size) and a recovery time of
at least $n(n-1)/2$ actions in total. We present a deterministic algorithm with
matching upper bounds that performs in arbitrary graphs. 
\item When probabilistic solutions are considered, we observe that at least 
$\Delta + 1$ states per process and a recovery time of $\Omega(n)$ actions in 
total are required. We present a probabilistically self-stabilizing algorithm 
that uses $\mathtt{k}$ states per process, where $\mathtt{k}$ is a parameter of the algorithm. 
When $k=\Delta+1$, the algorithm recovers in expected $O(\Delta n)$ actions. 
When $\mathtt{k}$ may grow arbitrarily, the algorithm recovers in expected $O(n)$ 
actions in total. Thus, our algorithm can be made optimal with respect to 
space or time complexity.
\end{enumerate}

\paragraph{Outline}

The remaining of the paper is organized as follows: Section~\ref{sec:model}
presents the programming model and problem specification,
Section~\ref{sec:impossible} provides impossibility results and lower bounds
for our problem, while Sections~\ref{sec:deterministic} and
\ref{sec:probabilistic} present matching upper bounds (in the deterministic
case) and asymptotically matching upper bounds (in the probabilistic case).
Section~\ref{sec:conclusion} gives some concluding remarks and open questions.

%% file: model.tex
\section{Model}
\label{sec:model}

\paragraph{Program model} 

A program consists of a set $V$ of $n$ processes. A process maintains a set of
variables that it can read or update, that define its \emph{state}. Each
variable ranges over a fixed domain of values. We use small case letters to
denote singleton variables, and capital ones to denote sets. A process contains
a set of \emph{constants} that it can read but not update. A binary relation
$E$ is defined over distinct processes such that $(i,j) \in E$ if and only if
$j$ can read the variables maintained by $i$; $i$ is a \emph{predecessor} of
$j$, and $j$ is a \emph{successor} of $i$. The set of predecessors (resp.
successors) of $i$ is denoted by $P.i$ (resp. $S.i$), and the union of
predecessors and successors of $i$ is denoted by $N.i$, the \emph{neighbors} of
$i$. In some case, we are interested in the iterated notions of those sets,
\emph{e.g.} $S.i^0 = {i}$, $S.i^1=S.i$, \ldots, $S.i^k=\cup_{j\in
S.i}S.j^{k-1}$. The values $\delta_{in}.i$, $\delta_{out}.i$, and $\delta.i$ denote
respectively $|P.i|$, $|S.i|$, and $|N.i|$; $\Delta_{in}$, $\Delta_{out}$,
and $\Delta$ denote the maximum possible values of $\delta_{in}.i$, $\delta_{out}.i$, and $\delta.i$ over all processes in $V$.

An action has the form $\langle name \rangle : \langle guard \rangle
\longrightarrow \langle command \rangle$. A \emph{guard} is a Boolean predicate
over the variables of the process and its communication neighbors. A
\emph{command} is a sequence of statements assigning new values to the
variables of the process. We refer to a variable $v$ and an action $a$ of
process $i$ as $v.i$ and $a.i$ respectively. A \emph{parameter} is used to
define a set of actions as one parameterized action. 

A \emph{configuration} of the program is the assignment of a value to every
variable of each process from its corresponding domain. Each process contains a
set of actions. An action is \emph{enabled} in some configuration if its guard
is \textbf{true} in this configuration. A \emph{computation} is a maximal
sequence of configurations such that for each configuration $\gamma_i$, the next
configuration $\gamma_{i+1}$ is obtained by executing the command of at least one
action that is enabled in $\gamma_i$. 
Maximality of a computation means that the
computation is infinite or it terminates in a configuration where none of the
actions are enabled. A program that only has terminating computations is 
\emph{silent}.

A \emph{scheduler} is a predicate on computations, that is, a scheduler is a
set of possible computations, such that every computation in this set satisfies
the scheduler predicate.  We distinguish two particular schedulers in the
sequel of the paper: the \emph{distributed} scheduler corresponds to predicate
\textbf{true} (that is, all computations are allowed); in contrast, the
\emph{locally central} scheduler implies that in any configuration belonging to
a computation satisfying the scheduler, no two enabled actions are executed
simultaneously on neighboring processes.

A configuration \emph{conforms} to a predicate if this predicate is
\textbf{true} in this configuration; otherwise the configuration \emph{violates} the
predicate. By this definition every configuration conforms to predicate
\textbf{true} and none conforms to \textbf{false}. Let $R$ and $S$ be
predicates over the configurations of the program. Predicate $R$ is
\emph{closed} with respect to the program actions if every configuration of the
computation that starts in a configuration conforming to $R$ also conforms to
$R$. Predicate $R$ \emph{converges} to $S$ if $R$ and $S$ are closed and any
computation starting from a configuration conforming to $R$ contains a
configuration conforming to $S$. The program \emph{deterministically
stabilizes} to $R$ if and only if \textbf{true} converges to $R$. The program
\emph{probabilistically stabilizes} to $R$ if and only if \textbf{true}
converges to $R$ with probability $1$.

\paragraph{Problem specification} 

Consider a set of colors ranging from $0$ to $\mathtt{k}-1$, for some integer
$\mathtt{k}\geq 1$. Each process $i$ defines a function $\mathit{color}.i$ that
takes as input the states of $i$ and its predecessors, and outputs a value in
$\{0,\ldots, \mathtt{k}-1\}$. The \emph{unidirectional vertex coloring}
predicate is satisfied if and only if for every $(i,j)\in E$, $\mathit{color}.i
\neq \mathit{color}.j$. 

%% file: impossibility.tex
\section{Impossibility results and lower bounds}
\label{sec:impossible}

\paragraph{General bounds}

We first observe two lower bounds that hold for any kind of silent program that
is self-stabilizing or probabilistically 
self-stabilizing for the unidirectional
coloring specification:

\begin{enumerate}
\item \emph{The minimal number of states per process is $\Delta+1$}. Consider a
bidirectional clique network (that is $(\Delta+1)$-sized), and a terminal
configuration of the program. Suppose that only $\Delta$ states are used, then
at least two processes $i$ and $j$ have the same state, and have the same view of
their predecessors. As a result $\mathit{color}.i = \mathit{color}.j$, and $i$
and $j$ being neighbors, the unidirectional coloring predicate does not hold in
this terminal configuration. 
\item \emph{The minimal number of moves overall is $\Omega(n)$}. Consider a
unidirectional chain of processes which are all initially in the same state.
For every process but one, the color is identical to that of its predecessor.
Since a change of state may only resolve two conflicts (that of the moving node
and that of its successor), a number of overall moves at least equal to
$\lfloor n/2 \rfloor$ is required, thus $\Omega(n)$ moves.  
\end{enumerate}

\paragraph{Deterministic bounds}

The rest of the section is dedicated to deterministic impossibility results and
lower bounds. Theorem~\ref{thm:impossibility_sync} (presented
below) shows that if unconstrained schedules (\emph{i.e.} the scheduler is distributed) 
are allowed, some initial
symmetric configurations can not be broken afterwards, making the
unidirectional coloring problem impossible to solve by a deterministic
algorithm. This justifies the later assumption of a locally central scheduler
in Section~\ref{sec:deterministic}, \emph{i.e.} a scheduler that never
schedules for execution two neighboring activatable processes \emph{simultaneously}.

\begin{lemma}
\label{lem:must_move}
Let an unidirectional network $\{p_0,p_1,\ldots,p_{n-1}\}$ of size $n$. 
Consider every node executes a uniform deterministic self-stabilizing uniform
coloring algorithm. Whenever there exists $i$ such that $s.p_i = s.p_{(i-1)
\bmod n}$, $p_i$ is activatable and if activated would change its state to
$s'.p_i$ with $s'.p_i \neq s.p_i$.
\end{lemma}

\begin{proof}
We first show that $p_i$ is activatable. Assume the contrary, and consider a
uniform cycle of $n$ nodes that are all in the same state. If $p_i$ is not
activatable, none of the remaining processes is activatable either. Hence, the
configuration is terminal. Now, a process $p_i$ may only read its own state and
that of its predecessor in the cycle, so the color $\mathit{color}.p_i$ is
uniquely determined by these two values only. Moreover, the
$\mathit{color}.p_i$ is the same as $\mathit{color}.p_{(i-1) \bmod n}$. So, the
configuration is terminal and two neighboring processes have the same color.
This contradicts the fact that the algorithm is a deterministic
self-stabilizing unidirectional coloring one.

Then we show that $p_i$, if activated moves to a different state $s'.p_i$.
Assume $p_i$ moves to the same state $s.p_i$, then if the starting
configuration is such that all nodes have the same state, then no node is able
to change its state, the algorithm being uniform. Since this configuration can
not be a coloring, the system never changes the global configuration and thus
is not self-stabilizing.
\end{proof} 

\begin{theorem}
\label{thm:impossibility_sync}
There exists no uniform deterministic self-stabilizing coloring algorithm that
can run on any unidirectional graph under a distributed scheduler.
\end{theorem}

\begin{figure}[P]
\centering
\subfigure[uniform configuration in state $s$]
{\includegraphics[scale=.35]{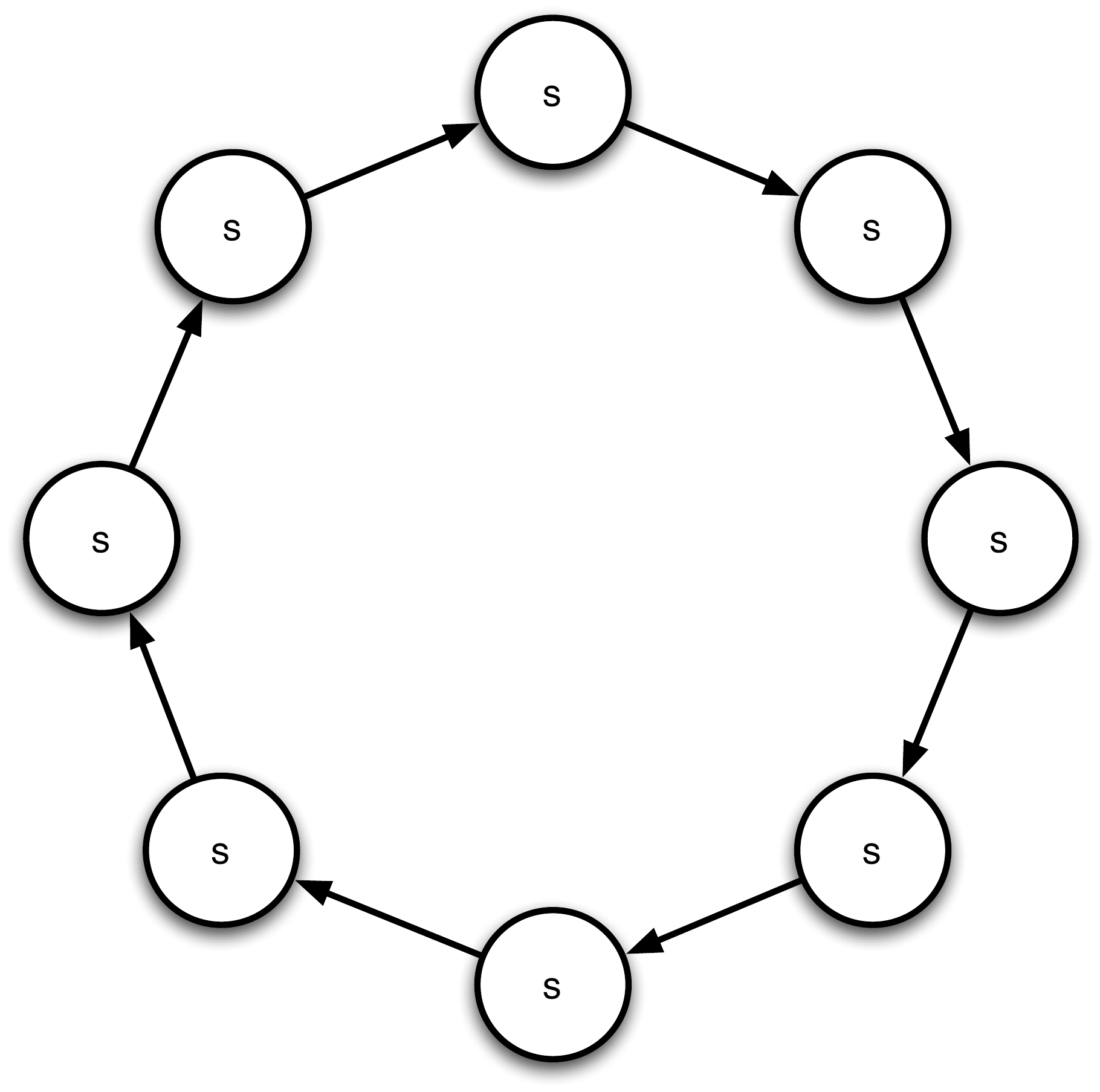}}
\subfigure[uniform configuration in state $s'$]
{\includegraphics[scale=.35]{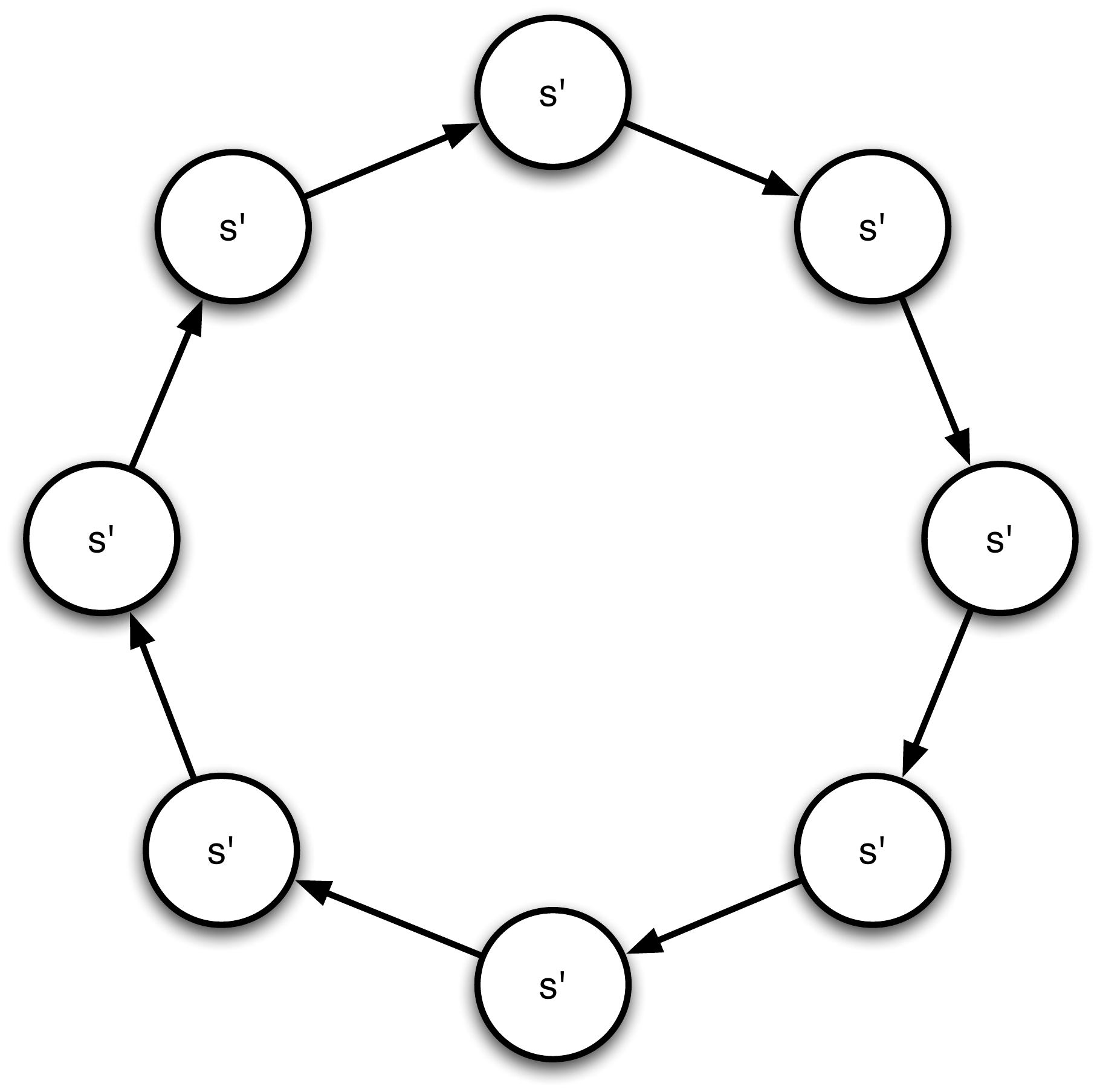}}
\caption{A possible execution with synchronous scheduling}
\label{fig:impossibility_sync}
\end{figure}

\begin{proof}
Assume the contrary. Consider a unidirectional cycle
$\{p_0,p_1,\ldots,p_{n-1}\}$ of size $n$. Assume that in the initial state all
nodes are in the same state $s$ (see Figure~\ref{fig:impossibility_sync}.(a)).
By Lemma~\ref{lem:must_move}, all nodes are activatable, and if a node is
activated by the scheduler, it moves to a different state $s'$. Consider the
synchronous scheduler that, at each step, activates all nodes. Then, after one
scheduler activation, all nodes have state $s'$ (see
Figure~\ref{fig:impossibility_sync}.(b)). After another activation, all nodes
move to state $s''$, etc. In this infinite execution, every configuration has
all nodes with the same state and thus, the coloration problem is not solved.
As a result, the algorithm can not be self-stabilizing.  
\end{proof}

Notice that the result of Theorem~\ref{thm:impossibility_sync} holds even if
the program is not required to be silent or if participating processes have
infinite number of states.

%% file: lower_bounds.tex
From now on, we assume the scheduler is locally central.
We demonstrate that a uniform silent deterministic
self-stabilizing algorithm for the unidirectional coloring problem must use at
least $n$ states per process in general networks. The proof is by exhibiting a
particular family of networks (namely, $n$-sized cycles) in which the bound is
reached by any such algorithm even assuming a locally central scheduler.
 
\begin{Algorithm}
\begin{tabbing}
12345\=12345\=12345\=12345\=\kill
\textbf{process} $i$\\
\textbf{const}\\
\> \texttt{k} : integer\\
\> $p.i$ : predecessor of $i$\\
\textbf{var}\\
\> $c.i$ : color of node $i$\\
\textbf{action}\\
\> $c.i = c.p.i$ $\rightarrow$\\
\>\> $c.i := c.i + 1 \mod \mathtt{k}$\\
\end{tabbing}
\caption{A uniform deterministic coloring algorithm for unidirectional rings}
\label{alg:deterministic}
\end{Algorithm}

\begin{lemma}
\label{lem:isomorphic}
Consider a unidirectional cycle $\{p_0,p_1,\ldots,p_{n-1}\}$ of size
$n$. Consider a scheduler that only activates a node $p_i$ 
when $s.p_i = s.p_{(i-1) \bmod n}$. Assume every node executes a uniform
deterministic self-stabilizing unidirectional coloring algorithm 
that uses a finite number of states $K$. There exists an initial
configuration such that the state sequence starting from this configuration 
is isomorphic to that of Algorithm~\ref{alg:deterministic}, 
for some parameter $k \leq K$.
\end{lemma}

\begin{figure}[P]
\centering
\subfigure[state transition function of a uniform deterministic 
self-stabilizing algorithm]{\includegraphics[scale=.35]{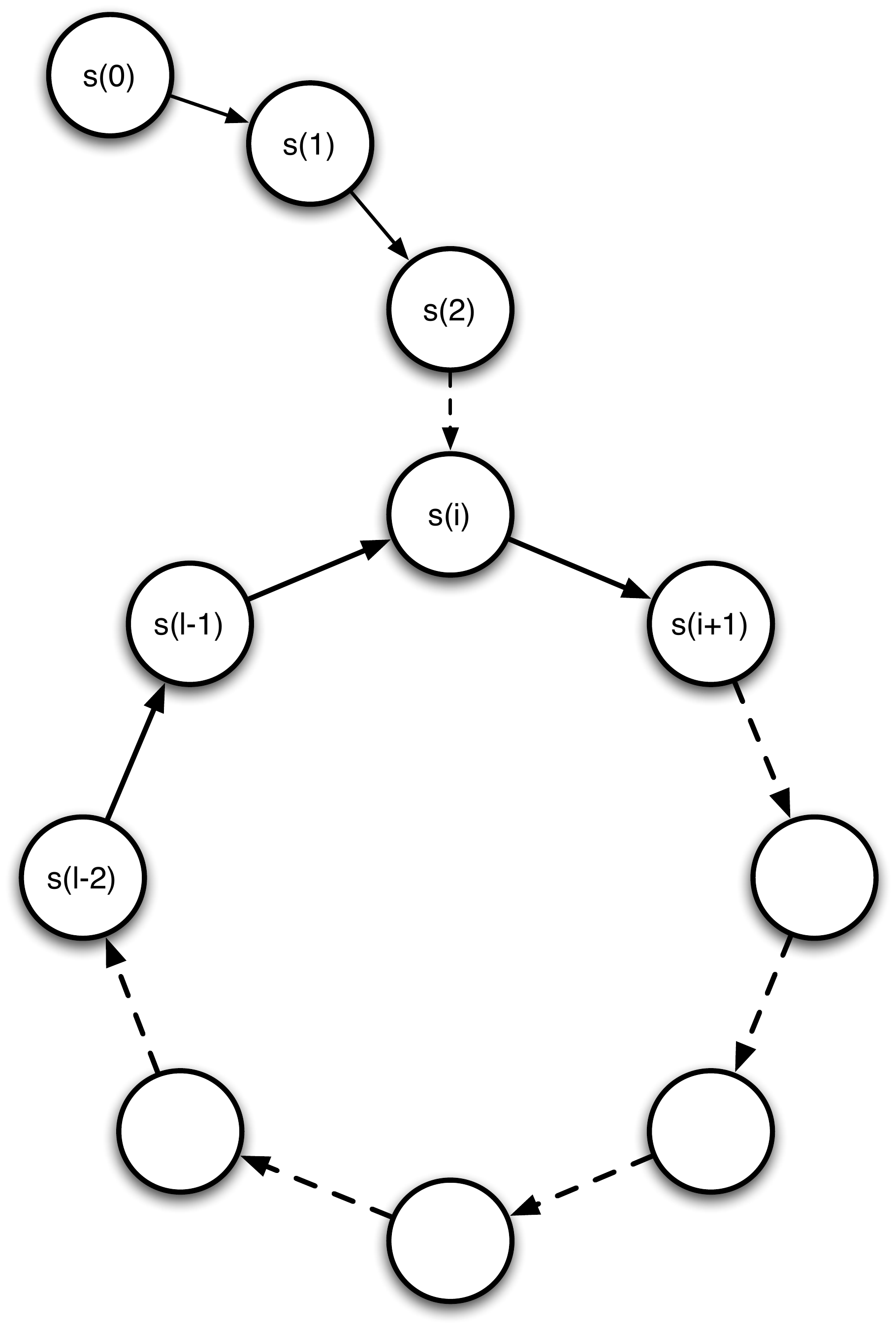}}
\subfigure[state transition function of 
Algorithm~\ref{alg:deterministic}]{\includegraphics[scale=.35]{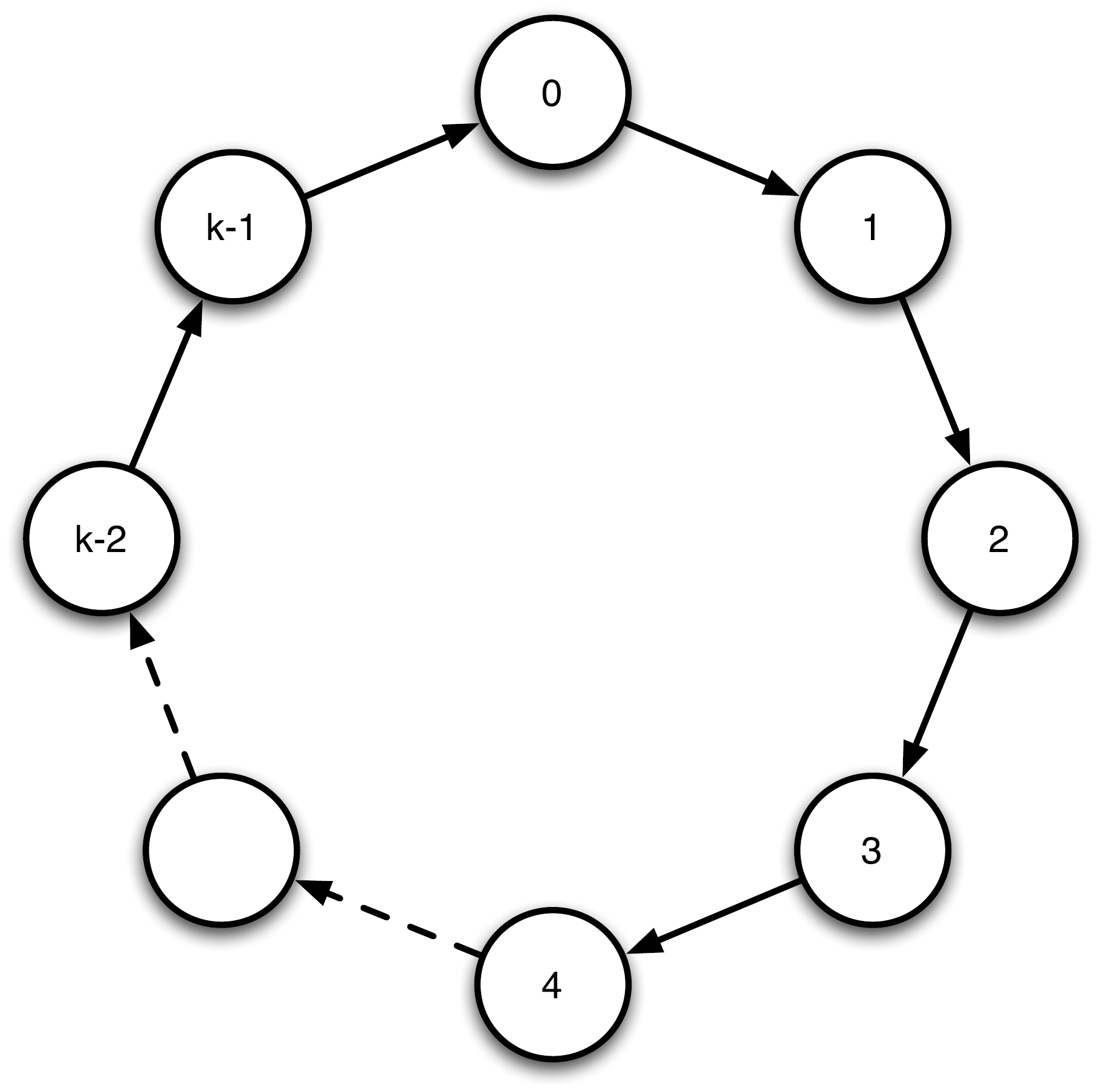}}
\caption{Isomorphism of the state transition function 
when $s.p_i = s.p_{i-1 \bmod n}$}
\label{fig:automata}
\end{figure}

\begin{proof} 
Consider a unidirectional cycle $\{p_0,p_1,\ldots,p_{n-1}\}$ of size $n$. 
 Assume a node $p_i$ is activated only when its state $s.p_i$ is equal to
$s.p_{(i-1) \bmod n}$, the state of the predecessor of $p_i$ in the cycle.
Then, the transition function of node $p_i$ is solely based on the state
$s.p_i$.  Let $s(0), s(1), \ldots  $ the sequence of states returned by the
transition function of process $p_i$ executing the self-stabilizing coloring
algorithm started in an arbitrary state $s(0)$. Note that \emph{(i)} the number
of states is finite, \emph{(ii)} a process with the same state as its
predecessor is always activatable (Lemma~\ref{lem:must_move}), and \emph{(iii)}
the protocol is deterministic. Then, the shape of the transition function of
$p_i$ is as depicted in Figure~\ref{fig:automata}.(a). That is, there exist $i$
and $l$ ($i<l$) such that $s(i)=s(l)$ and $l-i \leq K$.  Since the protocol is
self-stabilizing, it may be started from any arbitrary state, and in particular
from state $s(i)$ (in Figure~\ref{fig:automata}.(a)). Let denote $s(j)$ by
$j-i$, $\forall i \leq j < l$.  The transition function of $p_i$ is isomorphic
to that of the same process executing Algorithm~\ref{alg:deterministic} (given
in Figure~\ref{fig:automata}(b)) and assuming $k=l-i$.
\end{proof}

\begin{theorem}
\label{thm:lower_bound}
A silent uniform deterministic self-stabilizing protocol for unidirectional 
coloring requires 
at least $n$ states per process in a $n$-sized network (with $n\geq 2$).
\end{theorem}

\begin{figure}[P]
\centering
\subfigure[starting configuration]{\includegraphics[scale=.35]{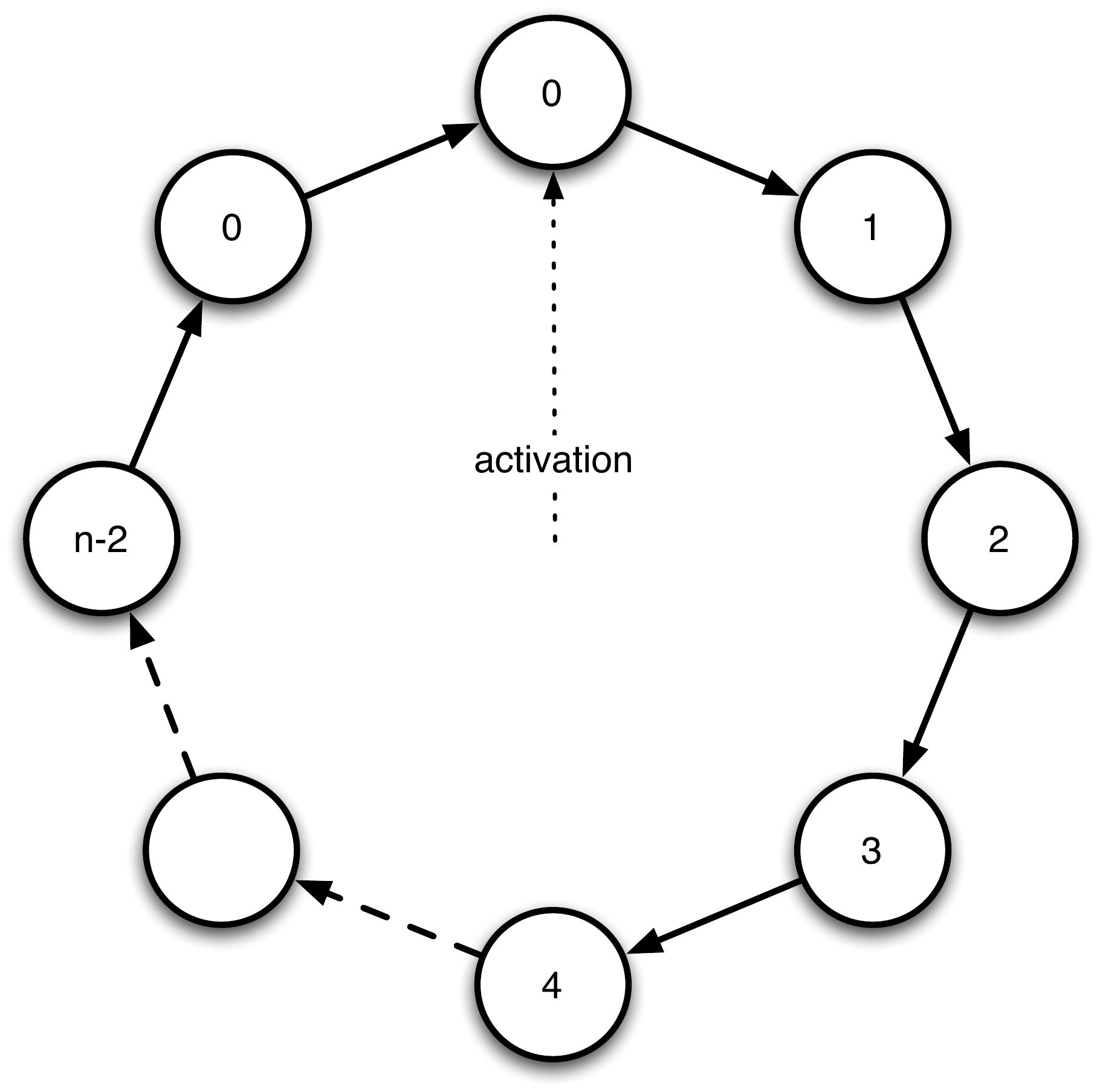}}
\subfigure[after $1$ step]{\includegraphics[scale=.35]{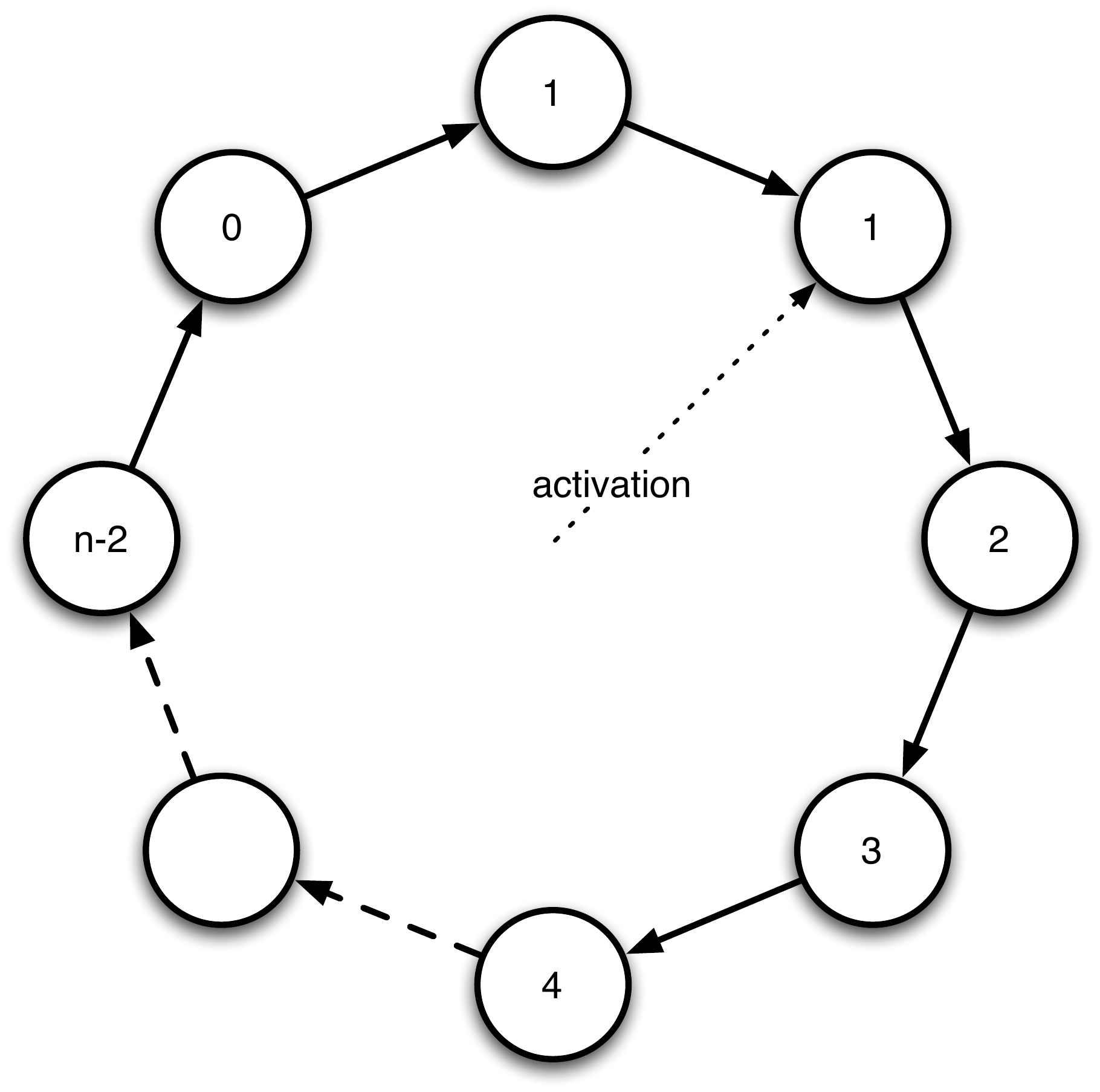}}
\subfigure[after $2$ steps]{\includegraphics[scale=.35]{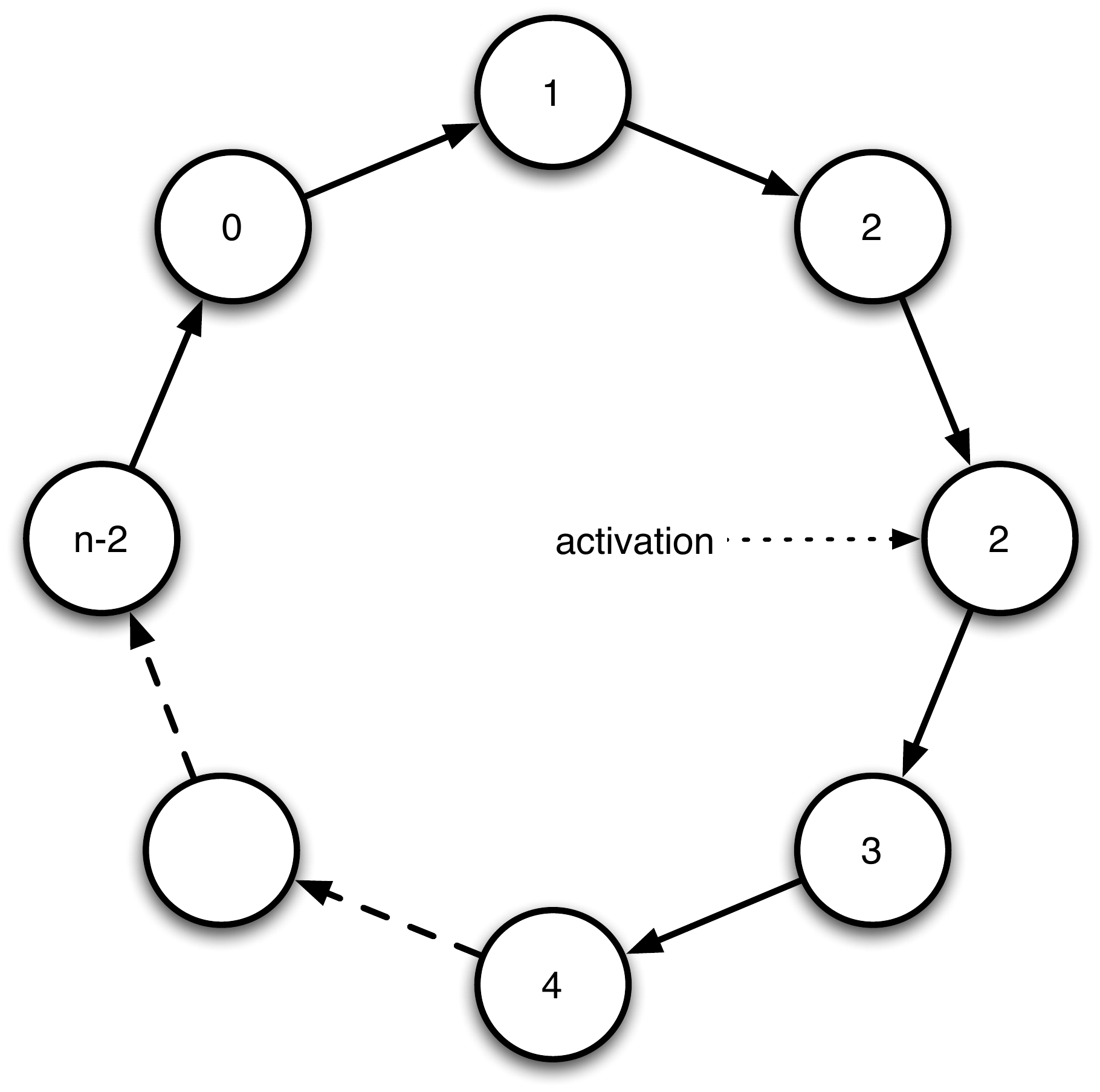}}
\subfigure[after $n-1$ steps]{\includegraphics[scale=.35]{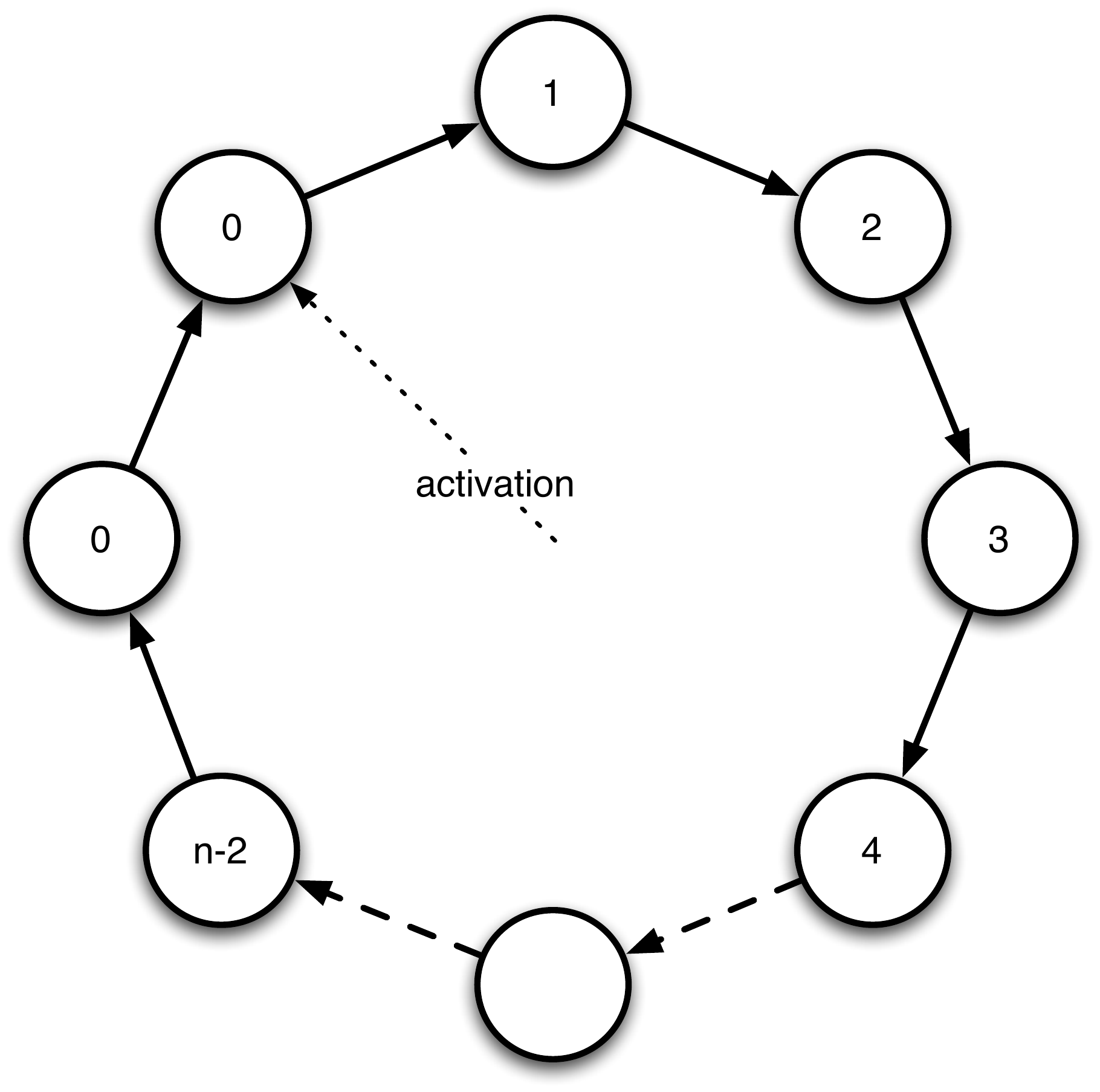}}
\caption{A possible execution of Algorithm~\ref{alg:deterministic}}
\label{fig:lower_bound}
\end{figure}

\begin{proof} 
Assume there exists a silent uniform deterministic
self-stabilizing protocol for coloring that requires less than $n$ states for a
particular node. Since the protocol is uniform, every node must use $k<n$
states. 

Consider a unidirectional cycle $\{p_0,p_1,\ldots,p_{n-1}\}$ of size $n$. In
what follows, we consider executions of the protocol in which the scheduler
only activates nodes that have the same state as their predecessor. By
Lemma~\ref{lem:isomorphic}, the transition function of every node is isomorphic
to that of Algorithm~\ref{alg:deterministic}, so we assume all nodes execute
Algorithm~\ref{alg:deterministic} with $k<n$.

In the following, we consider $k=n-1$ but the proof is easily expendable to any
$k<n$ by putting the $n-k+1$ last processors in the same state. Now consider
the unidirectional ring presented in Figure~\ref{fig:lower_bound}.(a). The
scheduler only activates the single node with the same state $0$ as its parent,
and reach the configuration presented in Figure~\ref{fig:lower_bound}.(b). The
scheduler may now activate the single node with the same state $1$ as its
parent and reach the configuration presented in
Figure~\ref{fig:lower_bound}.(c). We repeat the argument and reach the
configuration presented in Figure~\ref{fig:lower_bound}.(d). This configuration
is symmetric to that of the configuration presented in
Figure~\ref{fig:lower_bound}.(a), so the process can repeat infinitely often.
As a result, the protocol is not silent.  \end{proof}

We now address the question of time lower bounds for deterministic
self-stabilizing programs for the unidirectional coloring problem.

\begin{theorem}
A silent uniform deterministic self-stabilizing protocol for unidirectional 
coloring converges in at least $\frac{n(n-1)}{2}$ steps in general graphs.
\end{theorem}

\begin{proof}
Consider a chain topology, and assume that processors are ordered from the sink
$p_1$ to the source $p_n$. Assume all processors are initially in the same
state (a self-stabilizing program may start from any arbitrary configuration).
We now consider a locally central scheduler that activates nodes according to 
the schedule presented in Schedule~\ref{schedule:chain}.
\begin{Schedule}
\begin{tabbing}
12345\=12345\=12345\=12345\=\kill
\textbf{var}\\
\> $i$,$j$: integer\\
\textbf{scheduler}\\
\> \textbf{for} $j$ \textbf{from} $n-1$ \textbf{to} $1$\\
\>\>  \textbf{for} $i$ \textbf{from} $1$ \textbf{to} $j$\\
\>\>\> \textbf{activate} $p_i$
\end{tabbing}
\caption{Our $\frac{n(n-1)}{2}$-steps scheduling in $n$-sized chains}
\label{schedule:chain}
\end{Schedule}

Schedule~\ref{schedule:chain} selects a single process at a time for execution,
thus it satisfies the locally central scheduler property. In addition, it only
selects for execution a process that has the same state as its predecessor (and
thus activatable by Lemma \ref{lem:must_move}): if $p_1$ to $p_k$ 
have the same state $s$ then $p_1$ to $p_{k-1}$ are activatable and if they 
are activated in ascending order, $p_1$ to $p_{k-1}$ will move to the same 
``next'' state $s'$ (see Figure~\ref{fig:automata}.(a)). So every process 
activation leads to an effective move and the total number of 
activations is $\sum_{i=1}^{n-1}i=\frac{n(n-1)}{2}$.

Finally, all executions of the protocol must be terminating (it is silent), 
and from an initial configuration where all processes have the same state, 
a locally central schedule (like Schedule~\ref{schedule:chain}) may leads 
to $n(n-1)/2$ steps at least before termination. Hence the result.
\end{proof}

%% file: deterministic.tex
\section{Self-stabilizing deterministic unidirectional coloring}
\label{sec:deterministic}

In this section we propose a time and space optimal silent self-stabilizing
deterministic algorithm for unidirectional coloring.  The algorithm is referred thereafter as Algorithm
\ref{alg:deterministic_general} and performs under the locally central
scheduler. The algorithm can be informally described as follows: each process
$i$ has an integer variable $c.i$ (that ranges from $0$ to $\mathtt{k}-1$,
where $\mathtt{k}$ is a parameter of the algorithm) that denotes its color;
whenever a node has the same color as one of its predecessors, it changes its
color to the next available color (using the classical total order on
integers). Here, the $\mathit{color}.i$ function simply returns the color
variable $c.i$ of $i$.

\begin{Algorithm}
\begin{tabbing}
12345\=12345\=12345\=12345\=\kill
\textbf{process} $i$\\
\textbf{const}\\
\> \texttt{k} : integer\\
\> $P.i$ : set of predecessors of $i$\\
\textbf{parameter}\\
\> $p$ : node in $P.i$\\
\textbf{var}\\
\> $c.i$ : color of node $i$\\
\textbf{action}\\
\> $p \in P.i$, $c.i = c.p$ $\rightarrow$\\
\>\> \textbf{do} $p \in P.i$, $c.i = c.p$ $\rightarrow$\\
\>\>\> $c.i := c.i + 1 \mod \mathtt{k}$\\
\>\> \textbf{od}
\end{tabbing}
\caption{A uniform deterministic coloring algorithm for general unidirectional
networks}
\label{alg:deterministic_general}
\end{Algorithm}

A configuration is \emph{legitimate} if, for every process $i$, and for every
predecessor $p\in P.i$, $c.i \neq c.p.i$. Obviously, a legitimate configuration
satisfies the unidirectional coloring predicate (assuming $\mathit{color}.i$
return $c.i$) and is terminal (all guarded commands are disabled). There
remains to show how fast the algorithm attains a legitimate configuration in
the worst case for every possible locally central schedule.

\begin{theorem}
Algorithm~\ref{alg:deterministic_general} is a (state-optimal)
uniform silent deterministic self-stabilizing protocol for coloring nodes in
unidirectional general networks of size $n$ (when $\mathtt{k}=n$), assuming a
locally central scheduler and converges in $\frac{n(n-1)}{2}$
steps to a legitimate configuration. 
\end{theorem}

\begin{proof}
Assume Algorithm~\ref{alg:deterministic_general} starts in an arbitrary 
initial configuration $c$. 
We now consider the table that lists, for every possible color (in the set
$\{0, \ldots, n-1\}$ since we assume $\mathtt{k}=n$), the processes that
currently have this color. An example of such a table is presented as
Table~\ref{table:statetable}, where processes $P_3$ and $P_2$ have color $0$,
process $P_{n-2}$ has color $n-1$, etc. This table is denoted in the sequel as
the \emph{color table}.

\begin{table}
\begin{center}
\begin{tabular}{|l|c|c|c|c|c|c|}
   \hline
   Processors & $P_3, P_2$ & & $P_{n-1}, P_1$  & & ... & $P_{n-2}$ \\
   States     & $0$ & $1$ & $2$ & $3$ & ... & $n-1$ \\
   \hline
\end{tabular}
\end{center}
\caption{An example of color table}
\label{table:statetable}
\end{table}

According to Algorithm~\ref{alg:deterministic_general} the
evolution of the color table follows two rules: 
\begin{enumerate} 
\item A cell containing one process can not become empty.  That is, a process
having a color not used by any other process in the system can not be
activated. In our algorithm, this is due to the fact that processes are
activatable only if they share their color with their predecessor.   
\item A process only moves to the right (in a cyclic manner) and can not jump
over an empty cell.  Indeed, when activated, a process chooses the first
``next'' (in the sense of the usual total order on integers) unconflictual
color hence the processes always move to the right. A process may move by
several positions, but never skips a free position (this would mean that a
process does not choose the ''next'' color although this color is not
conflicting with any other process and thus not with the process
predecessors). 
\end{enumerate}

Since there are $n$ cells and $n$ processes, every process could be placed
in a different cell if necessary. Since a process can not jump over an empty
cell, after $n-1$ moves, a process is sure to find a free cell. In fact, the
number of moves a process may have to perform to reach a free cell depends on
the number of free cells. With $k$ free cells, there are at most $n-k$
consecutive non-empty cells that could potentially provoke further conflicts. A
process, in order to reach a free cell has to perform at most $n-k$ moves. Once
this process occupies a free cell, the number of free cells decreases to $k-1$.
Starting with $n-1$ free cell (every process has the same color), and finishing
with $1$, at most $1+2+...+(n-1)=\frac{n(n-1)}{2}$ steps are needed to have
every process in a free cell or with a non-conflicting color  (\emph{i.e.}
different from that of its predecessors) and thus reach a legitimate
configuration.
\end{proof}

%% file: probabilistic.tex
\section{Probabilistic self-stabilizing unidirectional coloring}
\label{sec:probabilistic}

In Section~\ref{sec:impossible}, we observed that there exist lowers bounds
even for probabilistic approaches to the unidirectional coloring problem. The
space lower bound is $\Delta+1$ and the time lower bound is $n$ (where $\Delta$
and $n$ are the degree and the size of the underlying simple undirected graph, 
respectively). 

The algorithm presented as Algorithm~\ref{alg:probabilistic_general}
can be informally described as follows. If a process has the same color as
one of its predecessors then it chooses a new color in the set of available
colors (\emph{i.e.} the set of colors that are not already used by any of its
predecessors). The colors are chosen in a set of size $\mathtt{k}$, where
$\mathtt{k}$ is a parameter of the algorithm.  In the following, we show that
Algorithm~\ref{alg:probabilistic_general} is probabilistically self-stabilizing
for the unidirectional coloring problem if $\mathtt{k} > \Delta$. To reach 
that goal we proceed in two steps: first we show that any terminal 
configuration satisfies the unidirectional coloring predicate 
(Lemma~\ref{theo:tc}); secondly, we show that the expected number of steps to 
reach a terminal configuration starting from an arbitrary one is bounded 
(Lemma~\ref{theo:limits}). 

\begin{Algorithm}
\begin{tabbing}
12345\=12345\=12345\=12345\=\kill
\textbf{process} $i$\\
\textbf{const}\\
\> \texttt{k} : integer\\
\> $P.i$ : set of predecessors of $i$\\
\> $C.i$ : set of colors of nodes in $P.i$\\
\textbf{parameter}\\
\> $p$ : node in $P.i$\\
\textbf{var}\\
\> $c.i$ : color of node $i$\\
\textbf{action}\\
\> $p \in P.i$, $c.i = c.p$ $\rightarrow$\\
\>\> $c.i := \mathtt{random}\left( \{0,\ldots ,\mathtt{k}-1\} 
\setminus C.i \right)$\\
\end{tabbing}
\caption{A uniform probabilistic coloring algorithm for general unidirectional
networks}
\label{alg:probabilistic_general}
\end{Algorithm}

\begin{lemma}
\label{theo:tc} 
Any terminal configuration satisfies the unidirectional coloring predicate.
\end{lemma} 

\begin{proof}
In a terminal configuration, every process $i$ satisfies $\forall j \in P.i,
c.i \neq c.j$ and $\forall j \in S.i, c.i \neq c.j$. Hence, in a terminal
configuration, every process $i$ has a color that is different from those of
its neighbors, which proves the theorem.  
\end{proof}

\begin{definition}[Conflict] 
Let $p$ be a process and $\gamma$ a configuration. The tuple
$(p$,$\gamma)$ is called a \emph{conflict} if and only if there exists
$q\in P.p$ (the predecessors of $p$) such that $c.q = c.p$ in $\gamma$.  
\end{definition}

\begin{lemma}\label{lem:strans}
Assume $\mathtt{k}>\Delta$. Let $(p$,$\gamma)$ be a conflict. The expected
number of conflicts created by the execution of one step of $p$ in order to
resolve $(p$,$\gamma)$ is:
\begin{equation} 
\frac{\delta.p - \delta_{in}.p}{\mathtt{k} - \delta_{in}.p}
\end{equation}
\end{lemma}

\begin{proof}
When a process $p$ executes Action $\A$ from $\gamma$, it chooses a
new color in a set of at least $\mathtt{k} - \delta_{in}.p$
colors. That is, there are  $\mathtt{k}$ colors and it can not choose a
color chosen by one of its predecessors, therefore at most $\delta_{in}.p$
colors are removed from the set of possible choices.

For each $q \in S.p \wedge q \notin P.p$, $p$ and $q$ are in conflict if and
only if, $p$ chooses the color of $q$. Notice that $p$ has $\frac{1}
{(\mathtt{k} - \delta_{in}.p)}$ chance to create a new conflict. Since the
number of successors of $p$ not in the set of predecessors of $p$, $\sharp \{q
\in S.p \wedge q \notin P.p\}$, is $\delta.p-\delta_{in}.p$, the expected
number of created conflicts is $\frac{\delta.p - \delta_{in}.p}{\mathtt{k} -
\delta_{in}.p}$. 
\end{proof}

\begin{lemma}\label{lem:generalisation}
Let $(p,\gamma)$ be a conflict. The expected number of conflicts created by the
execution of one step of $p$ in order to resolve $(p$,$\gamma)$ is less than or
equal to:
\begin{equation} 
\M = \frac{\Delta - 1}{\mathtt{k} - 1}, ~\mathtt{k}>\Delta
\end{equation}
\end{lemma}

\begin{proof} 
Observe that $\forall p \in V, \Delta \geq \delta.p$, therefore
$\forall p \in V, \frac{\delta.p - \delta_{in}.p}{\mathtt{k} - \delta_{in}.p}
\leq \frac{\Delta - \delta_{in}.p}{\mathtt{k} - \delta_{in}.p}$. In order to
find an upper bound for this value, let $f:p \in V, \Delta<\mathtt{k},
\delta_{in}.p\in[0,\Delta], \delta_{in}.p \mapsto
\frac{\Delta-\delta_{in}.p}{\mathtt{k}-\delta_{in}.p}$.  Its derivative exists
and is
$f^\prime(\delta_{in}.p)=\frac{\Delta-\mathtt{k}}{(\mathtt{k}-\delta_{in}.p)^2}$.
By hypothesis, $\mathtt{k} > \Delta$, so $f^\prime(\delta_{in}.p)<0$ and $f$ is
decreasing. Therefore, $f(\delta_{in}.p)$ is maximum when $\delta_{in}.p=0$ but
for this value $(p$,$\gamma)$ can not be a conflict. Therefore, $\delta_{in}.p
\geq 1$ and $f$ is maximum for $\delta_{in}.p = 1$ which leads to
$\frac{\delta.p - \delta_{in}.p}{\mathtt{k} - \delta_{in}.p} \leq \frac{\Delta
- \delta_{in}.p}{\mathtt{k} - \delta_{in}.p} \leq \frac{\Delta - 1}{\mathtt{k}
- 1}$. 
\end{proof}

\begin{lemma}\label{lem:initial}
Let $(p,\gamma)$ be a conflict. The expected
number of step created in 
order to resolve this conflict is less than:
\begin{equation} 
\frac{\mathtt{k} - 1}{\mathtt{k}-\Delta}, ~\mathtt{k}>\Delta  
\end{equation}
\end{lemma}

\begin{proof}
From Lemma \ref{lem:generalisation}, the expected number of conflicts
created by one step of 
$p$ is less than $\M=\frac{\Delta - 1}{\mathtt{k} - 1}$. 
Then, the processes in $S.p$ who received the created conflicts
produce at most $\M$ new conflicts each since there are at most 
$\M$ expected such processes. They will perform $\M$ steps and
create at most $\M^2=\left(\frac{\Delta - 1}{\mathtt{k} - 1}\right)^2$
new conflicts. Then the $\M^2$ processes in $S.p^2$ (the successors at
distance two from $p$), after $\M^2$ step (one for each) will produce
$\M^3$ new conflicts. By recurrence, at most $\M^i$ expected steps 
will be executed and $\M^{i+1}$ new conflicts will be created by the
processes in $S.p^i$ (the successors at distance $i$ from $p$).

Finally, the expected number of steps executed is
$\sum^\infty_{i=0}{\M^i}=\sum^\infty_{i=0}{\left(\frac{\Delta - 1}{\mathtt{k} -
1}\right)^i}$. Note that $\Delta \geq 1$ and $\mathtt{k}>\Delta$, so $0 <
\frac{\Delta - 1}{\mathtt{k} - 1} < 1$. The expected number of steps executed
in order to solve the conflict $(p,\gamma)$ is
$\sum^\infty_{i=0}{\left(\frac{\Delta - 1}{\mathtt{k} - 1}\right)^i} =
\frac{1}{1 - \frac{\Delta - 1}{\mathtt{k} - 1}} = \frac{\mathtt{k} -
1}{\mathtt{k}-\Delta}$
\end{proof}

\begin{lemma}\label{theo:limits}
Starting from an arbitrary configuration,
the expected number of steps to reach a configuration verifying the
unidirectional 
coloring predicate is less or equal to:
\begin{equation}
\frac{n(\mathtt{k}-1)}{\mathtt{k}-\Delta}, ~\mathtt{k}>\Delta
\end{equation}
\end{lemma}

\begin{proof}
In the worst case the number of initial conflicts is $n$. Then the proof is a 
direct consequence of Lemma \ref{lem:initial}.
\end{proof}

Notice that with a minimal number of colors (\emph{i.e.}, $\mathtt{k} = \Delta
+ 1$), the expected number of steps to reach a terminal configuration 
starting from an arbitrary configuration is less than $n\Delta$.
Moreover, when the number of colors increases (\emph{i.e.}, $\mathtt{k}
\rightarrow \infty$), 
the expected number of steps to reach a terminal configuration
starting from an arbitrary configuration converges to $n$.

\begin{theorem}
Algorithm \ref{alg:probabilistic_general} is a probabilistic self-stabilizing
solution for the unidirectional coloring when $\mathtt{k} > \Delta$.
\end{theorem}

\begin{proof}
The proof is a direct consequence of Lemma~\ref{theo:tc} and Lemma~\ref{theo:limits}.
\end{proof}

%% file: conclusion.tex
\section{Conclusion}
\label{sec:conclusion}
We investigated the intrinsic complexity of performing local tasks in
unidirectional networks in a self-stabilizing setting. Contrary to
``classical'' bidirectional networks, local vertex coloring now induces global
complexity ($n$ states per process at least, $n$ moves per process at least)
for deterministic solutions. We presented state and time optimal solutions for
the deterministic case, and asymptotically optimal solutions for the
probabilistic case. This work raises several important open questions:

\begin{enumerate}
\item Our probabilistic solution can be tuned to be optimal in space (and is then with a
$\Delta$ multiplicative penalty in time), or optimal in time, but not both.
However, our lower bounds do not preclude the existence of probabilistic
solutions that are optimal for both complexity measures.
\item Several of the lower bounds we provide in the deterministic case rely on
the silence property of the expected solution. We question the possibility of
designing deterministic algorithms that are not silent yet provide coloring
with less than $n$ colors in general graphs.
\end{enumerate}

%% file: RR-6524.bbl
\begin{thebibliography}{10}

\bibitem{AB98j}
Yehuda Afek and Anat Bremler-Barr.
\newblock Self-stabilizing unidirectional network algorithms by power supply.
\newblock {\em Chicago J. Theor. Comput. Sci.}, 1998, 1998.

\bibitem{AD02j}
Yehuda Afek and Shlomi Dolev.
\newblock Local stabilizer.
\newblock {\em J. Parallel Distrib. Comput.}, 62(5):745--765, 2002.

\bibitem{BDDT07j}
Joffroy Beauquier, Sylvie Dela\"{e}t, Shlomi Dolev, and S\'{e}bastien Tixeuil.
\newblock Transient fault detectors.
\newblock {\em Distributed Computing}, 20(1):39--51, 2007.

\bibitem{BGJ07j}
Joffroy Beauquier, Maria Gradinariu, and Colette Johnen.
\newblock Randomized self-stabilizing and space optimal leader election under
  arbitrary scheduler on rings.
\newblock {\em Distributed Computing}, 20(1):75--93, 2007.

\bibitem{CG01c}
Jorge~Arturo Cobb and Mohamed~G. Gouda.
\newblock Stabilization of routing in directed networks.
\newblock In Datta and Herman \cite{DBLP:conf/wss/2001}, pages 51--66.

\bibitem{DDT99j}
Sajal~K. Das, Ajoy~Kumar Datta, and S\'{e}bastien Tixeuil.
\newblock Self-stabilizing algorithms in dag structured networks.
\newblock {\em Parallel Processing Letters}, 9(4):563--574, December 1999.

\bibitem{DBLP:conf/wss/2001}
Ajoy~Kumar Datta and Ted Herman, editors.
\newblock {\em Self-Stabilizing Systems, 5th International Workshop, WSS 2001,
  Lisbon, Portugal, October 1-2, 2001, Proceedings}, volume 2194 of {\em
  Lecture Notes in Computer Science}. Springer, 2001.

\bibitem{DDT06j}
Sylvie Dela\"{e}t, Bertrand Ducourthial, and S\'{e}bastien Tixeuil.
\newblock Self-stabilization with r-operators revisited.
\newblock {\em Journal of Aerospace Computing, Information, and Communication},
  2006.

\bibitem{D74j}
Edsger~W. Dijkstra.
\newblock Self-stabilizing systems in spite of distributed control.
\newblock {\em Commun. ACM}, 17(11):643--644, 1974.

\bibitem{D00b}
S.~Dolev.
\newblock {\em Self-stabilization}.
\newblock MIT Press, March 2000.

\bibitem{DGS99j}
Shlomi Dolev, Mohamed~G. Gouda, and Marco Schneider.
\newblock Memory requirements for silent stabilization.
\newblock {\em Acta Inf.}, 36(6):447--462, 1999.

\bibitem{DIM97j}
Shlomi Dolev, Amos Israeli, and Shlomo Moran.
\newblock Resource bounds for self-stabilizing message-driven protocols.
\newblock {\em SIAM J. Comput.}, 26(1):273--290, 1997.

\bibitem{DS04j}
Shlomi Dolev and Elad Schiller.
\newblock Self-stabilizing group communication in directed networks.
\newblock {\em Acta Inf.}, 40(9):609--636, 2004.

\bibitem{DHT04ca}
Philippe Duchon, Nicolas Hanusse, and S\'{e}bastien Tixeuil.
\newblock Optimal randomized self-stabilizing mutual exclusion in synchronous
  rings.
\newblock In {\em Proceedings of the 18th Symposium on Distributed Computing
  (DISC 2004)}, number 3274 in Lecture Notes in Computer Science, pages
  216--229, Amsterdam, The Nederlands, October 2004. Springer Verlag.

\bibitem{DT01jb}
Bertrand Ducourthial and S\'{e}bastien Tixeuil.
\newblock Self-stabilization with r-operators.
\newblock {\em Distributed Computing}, 14(3):147--162, July 2001.

\bibitem{DT03j}
Bertrand Ducourthial and S\'{e}bastien Tixeuil.
\newblock Self-stabilization with path algebra.
\newblock {\em Theoretical Computer Science}, 293(1):219--236, 2003.
\newblock Extended abstract in Sirrocco 2000.

\bibitem{GT02c}
Christophe Genolini and S\'{e}bastien Tixeuil.
\newblock A lower bound on k-stabilization in asynchronous systems.
\newblock In {\em Proceedings of IEEE 21st Symposium on Reliable Distributed
  Systems (SRDS'2002)}, Osaka, Japan, October 2002.

\bibitem{GT00c}
Maria Gradinariu and S\'{e}bastien Tixeuil.
\newblock Self-stabilizing vertex coloring of arbitrary graphs.
\newblock In {\em International Conference on Principles of Distributed Systems
  ({OPODIS}'2000)}, pages 55--70, Paris, France, December 2000.

\bibitem{MT07j}
Toshimitsu Masuzawa and S\'{e}bastien Tixeuil.
\newblock Stabilizing link-coloration of arbitrary networks with unbounded
  byzantine faults.
\newblock {\em International Journal of Principles and Applications of
  Information Science and Technology (PAIST)}, 1(1):1--13, December 2007.

\bibitem{MFGST06c}
Nathalie Mitton, Eric Fleury, Isabelle Gu\'{e}rin-Lassous, Bruno S\'{e}ricola,
  and S\'{e}bastien Tixeuil.
\newblock On fast randomized colorings in sensor networks.
\newblock In {\em Proceedings of ICPADS 2006}, pages 31--38. IEEE Press, July
  2006.

\bibitem{NA02c}
Mikhail Nesterenko and Anish Arora.
\newblock Tolerance to unbounded byzantine faults.
\newblock In {\em 21st Symposium on Reliable Distributed Systems (SRDS 2002)},
  pages 22--. IEEE Computer Society, 2002.

\bibitem{T01c}
S{\'e}bastien Tixeuil.
\newblock On a space-optimal distributed traversal algorithm.
\newblock In Datta and Herman \cite{DBLP:conf/wss/2001}, pages 216--228.

\end{thebibliography}
